\documentclass[sigconf]{acmart}
\usepackage{colortbl}
\usepackage{makecell}

\usepackage{amssymb}
\usepackage{makecell}
\usepackage{natbib}
\usepackage{datatool}
\usepackage{etoolbox}
\DTLnewdb{bibnotes}

\makeatletter
\patchcmd{\@lbibitem}%
  {\item[\hfil\NAT@anchor{#2}{\NAT@num}]}%
  {%
    \item[\hfil\NAT@anchor{#2}{\NAT@num}]%
    \DTLforeach[\DTLiseq{\mylabel}{#2}]{bibnotes}{\mylabel=mylabel,\mynote=mynote}{\textit{\mynote}}
  }{}{\message{^^JPatching failed^^J}}%

\makeatother

\AtBeginDocument{%
  \providecommand\BibTeX{{%
    \normalfont B\kern-0.5em{\scshape i\kern-0.25em b}\kern-0.8em\TeX}}}

\acmConference[Conference acronym 'XX]{Make sure to enter the correct
  conference title from your rights confirmation email}{June 03--05,
  2018}{Woodstock, NY}
\acmPrice{15.00}
\acmISBN{978-1-4503-XXXX-X/18/06}

\setcopyright{acmcopyright}
\copyrightyear{2022}
\acmYear{2022}
\acmDOI{10.1145/1122445.1122456}

\acmConference[Chinese CHI '22]{Chinese CHI '22: The Tenth International Symposium of Chinese CHI (Chinese CHI 2022)}{October 23--October 26, 2022}{Guangzhou, China}
\acmBooktitle{Chinese CHI '22: The Tenth International Symposium of Chinese CHI (Chinese CHI 2022),
  October 22--October 23, 2022, Guangzhou, China}
\acmPrice{15.00}
\acmISBN{978-1-4503-XXXX-X/18/06}

\begin{document}

\title[Communication in Immersive Social Virtual Reality]{Communication in Immersive Social Virtual Reality: A Systematic Review of 10 Years' Studies}



\author{Xiaoying Wei}
\authornote{Both authors contributed equally to this research.}
\affiliation{
  \institution{The Hong Kong University of Science and Technology}
  \city{Hong Kong SAR}
  \country{China}
}
\email{xweias@connect.ust.hk}

\author{Xiaofu Jin}
\authornotemark[1]
\affiliation{
  \institution{The Hong Kong University of Science and Technology}
  \city{Hong Kong SAR}
  \country{China}
}
\email{xjinao@connect.ust.hk}

\author{Mingming Fan}
\authornote{Corresponding Author}
\orcid{0000-0002-0356-4712}
\affiliation{
 \institution{The Hong Kong University of Science and Technology (Guangzhou)}
  \city{Guangzhou}
  \country{China}
}
\affiliation{
  \institution{The Hong Kong University of Science and Technology}
  \city{Hong Kong SAR}
  \country{China}
}
\email{mingmingfan@ust.hk}




\renewcommand{\shortauthors}{Wei, Jin and Fan}

\begin{abstract}
  As virtual reality (VR) technologies have improved in the past decade, more research has investigated how they could support more effective communication in various contexts to improve collaboration and social connectedness. However, there was no literature to summarize the uniqueness VR provided and put forward guidance for designing social VR applications for better communication. To understand how VR has been designed and used to facilitate communication in different contexts, we conducted a systematic review of the studies investigating communication in social VR in the past ten years by following the PRISMA guidelines. We highlight current practices and challenges and identify research opportunities to improve the design of social VR to better support communication and make social VR more accessible.  
\end{abstract}



\maketitle
\section{Introduction}

 Communication is defined as the transmission of resources, such as knowledge, data, and skills, among different parties using shared symbols and media \cite{cheng2001network, wen2020using}. Efficient communication is an essential demand which would impact work efficiency and user experiences in contexts such as collaboration \cite{zhang2012social}, social contact \cite{Maloney2020}, meeting \cite{Abdullah2021}, education \cite{riemer2007communication}, and gaming \cite{zhao2018hand}. Recently, with the outbreak of COVID-19, the need for remote communication has increased dramatically \cite{ERSEK2021213,shockley2021remote, Heshmat2021Family, Lei2022Emergency}. The current mainstream technology for remote communication is video conferencing systems such as Zoom and Skype. However, these types of communication media do not provide an immersive environment as face-to-face does. As a result, they do not provide users with the feeling of bodily closeness, emotional closeness, and the experiences of physical presence. They also limit users' interaction with the environment and objects, such as visiting museums or traveling together \cite{fuchsberger2021grandparents}.

 Virtual reality (VR) refers to a process of mental transcendence into synthetic, three-dimensional (3D) virtual environments with the use of immersive technologies \cite{1994What, zhao2009survey}. The concept of a virtual environment facilitating communication has been suggested and analyzed in the 1990s \cite{biocca1992communication, biocca1995virtual}. With the immersive environment and the representation of the avatar, VR can provide users with real-time and embodied interactions that are similar to face-to-face communication rather than merely looking at a computer screen. It also affords a broader spectrum of communication modes including both verbal and non-verbal interactions such as voice, gestures, proximity, facial expression and haptic feedback. Multiple users can interact with one another via VR head-mounted displays (HMDs), which is known as \textit{social VR}. In the past decade, booming commercial social VR applications such as Facebook Spaces, VR Chat, AltspaceVR, and Rec Room have led to an emerging research agenda on social VR in HCI and CSCW, drawing research attention to new research questions, especially in communication in social VR environments \cite{Li2019, Baker2019, Abdullah2021, Smith2018}. 
 
 Although there are various applications of social VR, the core of social VR is to provide communication between people as a bridge. Prior research explored communication in social VR in different aspects such as the types of non-verbal cues \cite{Maloney2020}, compared with video conferencing tools \cite{Abdullah2021}. While informative, there still lacks a holistic understanding of how communication has been studied in immersive social VR. This has motivated us to conduct a systematic literature review to better understand what has been studied in the past decades of literature regarding communication in social VR and to highlight the design opportunities for future work. Specifically, we aim to answer the following two research questions:   
 
 \begin{enumerate}

 \item   RQ1: What factors in social VR affect users' communication quality? What are the differences between social VR and other forms of media for communication?

 \item   RQ2: What methods are used to evaluate communication quality in social VR? 

 \end{enumerate}
 
We conducted a systematic review using the widely adopted PRISMA method \cite{moher2009preferred} to investigate communication in social VR. We focused on the relevant papers published in major human-computer interaction venues (e.g., CHI, TOCHI, UIST, VR, TVCG, DIS) in the past ten years (2012-2022), when VR-related research has gained increasing attention from both academia and industry. By answering the two RQs through this literature review, we make the following two contributions.
 
\begin{enumerate}
 \item We identified the factors that affect people's communication quality in social VR, including the sense of anonymity brought by the avatar, a diverse set of approaches to expressing information including natural expressions and actions with the representation by the avatar. 
 \item We investigated the methods used to study communication in social VR in the reviewed papers and identified research opportunities to better support communication and make it more accessible in social VR. We propose future research directions according to the results of our review, such as using social VR as communication media for remote family relatives.
\end{enumerate}

\section{Background and Related Work}

\subsection{Social VR}

 Social VR refers to the applications (apps) which enable people to interact with each other through virtual environments via VR head-mounted displays (HMDs) \cite{li2021social}. Recently, commercial social VR applications attracted many users to interact with one another, such as VRChat, Rec Room, AltspaceVR, High Fidelity, Facebook Spaces, and so on. Previous research has explored the design guidelines and user experiences in these commercial social VR apps \cite{Maloney2020, Tanenbaum2020}. In addition to commercial social VR apps, researchers have also explored social VR in different aspects. For example, prior research investigated collaborative virtual environments (CVEs) \cite{churchill2012collaborative, benford2001collaborative}, ethnographies of virtual worlds \cite{taylor2009play, boellstorff2012ethnography}, and related accounts of social activity in virtual environments \cite{castronova2008exodus, dibbell1998my, ondrejka2004piece}. No matter how various applications of social VR is, the core of social VR is to provide communication between people as a bridge. In this paper, we focus on exploring communication in social VR.

\subsection{Communication in social VR}

 Communication is an interactive process through which participants mutually exchange and interpret verbal and nonverbal messages \cite{morency2010modeling}. Communication is important, acting as a strong need and desire for people with their distant family and friends. \cite{neustaedter2006interpersonal, tee2009exploring, romero2007connecting}. 
 Moreover, good communication helps people build trust, make good relationships with one another, solve problems and handle conflicts \cite{communicationBenefits}. To facilitate communication, social media apps such as Zoom and Facebook are widely used around the world, enabling people to experience remote communication using text, audio, or video.
 
 However, these types of communication media do not provide an immersive communication environment compared with face-to-face communication, making people feel less present and lack bodily closeness, emotional closeness, as well as opportunities to interact with the physical environment and objects together \cite{fuchsberger2021grandparents}. With VR, an emerging alternative, users are able to "meet" in a shared, immersive virtual environment and interact with virtual representations of each other, thus bringing a better communication experiences \cite{heidicker2017influence, Li2019}. Researchers have already explored communication in social VR. Maloney et al. explored the types of non-verbal interactions used naturally in social VR and participants' perceptions of non-verbal communication as well as the resulting interaction outcome, then highlighted potential design implications that aim at better supporting non-verbal communication in social VR \cite{Maloney2020}. Abdullah's work compared people’s behavioral patterns across the VR and videoconference \cite{Abdullah2021}, and observed significant behavioral differences. Although informative, there is no literature to summarize what capabilities VR provides to make users' communication in VR different from other media. In addition, there is no summarization of how researchers measure the communication effect and user experiences in different contexts in social VR, which makes it hard for researchers to refer to. Therefore, we conducted a systematic literature review to address this gap, which can help researchers and designers to design more meaningful and practical VR applications for communication.

\section{Method}

\begin{figure}[]
\includegraphics[width=3.7 in]{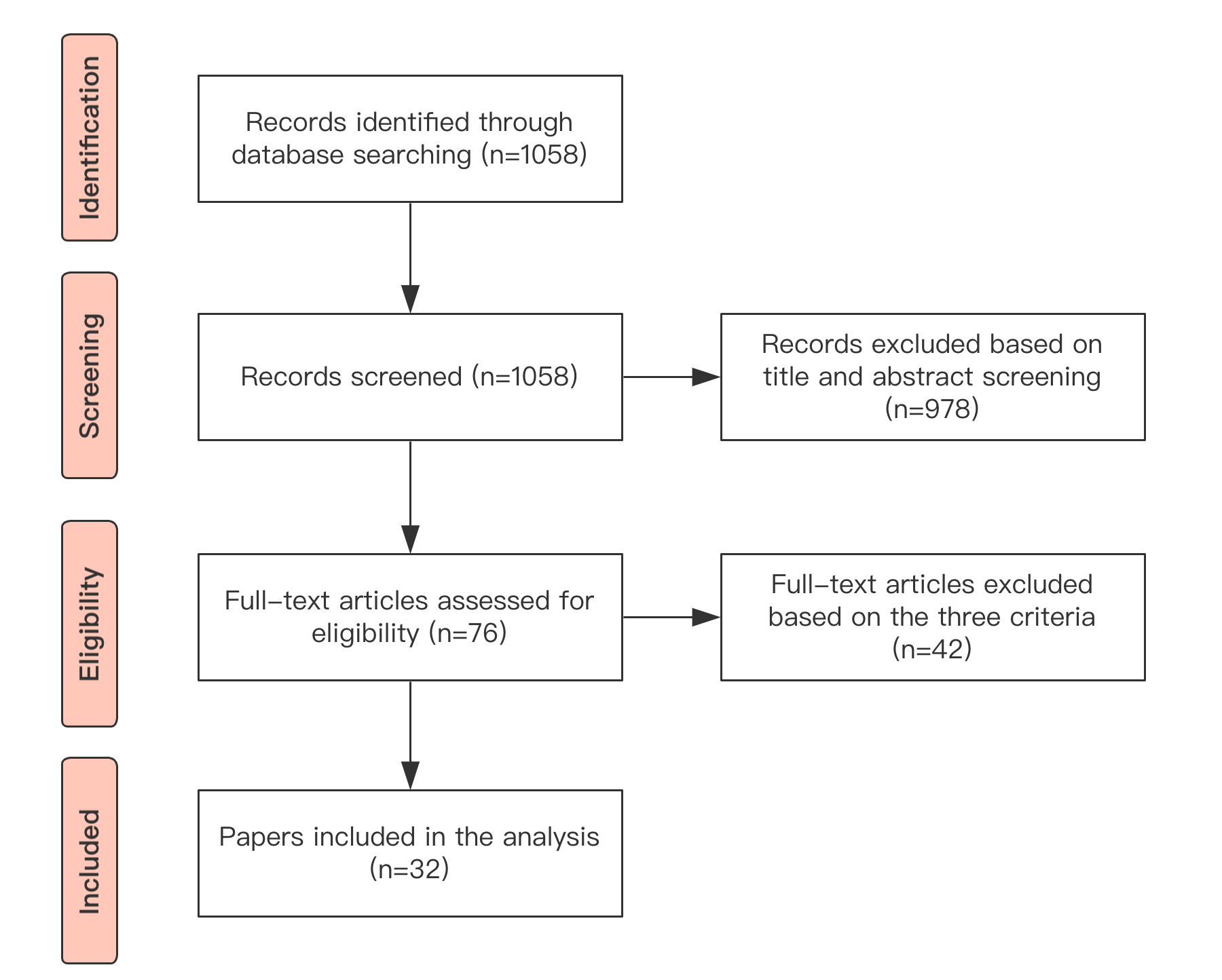}
\vspace{-0.2cm}
\caption{Our literature search and inclusion phases used the PRISMA procedure. This diagram describes the information flow throughout the following four phases, identification, screening, eligibility, and inclusion, and presents the number of the literature identified, included, excluded, and the reasons for exclusions.} 
\label{fig:process}
\end{figure}

 We conducted the systematic literature review following the PRISMA method \cite{moher2009preferred}. PRISMA is a widely used method to ensure the reproducibility of the literature review in many disciplines, including human-computer interaction~\cite{bergstrom2021evaluate, tuena2020usability, macarthur2021you}. The PRISMA method contains four phases to identify eligible papers. Figure ~\ref{fig:process} shows the details of the four phases in our research. The following section explains how we conducted each phase in detail.

\subsection{Identification}

 We chose ACM Digital Library, IEEE Xplore, and Springer as our targeted databases because these databases publish immersive VR-related papers from significant HCI conferences and journals (e.g., CHI, TOCHI, UIST, VR, TVCG, DIS). We also conducted a supplementary search using the relevant keywords in Google Scholar to avoid omissions.

 Table ~\ref{tab:Boolean} shows our search query in three databases. Taking the search query of ACM Digital Library as an example, to understand how users communicate in social VR, we included 'communicat* and (virtual or VR)' in our search query and allowed them to appear anywhere in the title or the abstract, and allowed 'social' to appear anywhere in the full text.

 \begin{table*}[htb!]
    \caption{Boolean instructions for ACM Digital Library, IEEE Xplore and Springer}
    \label{tab:Boolean}
    \renewcommand{\arraystretch}{1.3}
    \begin{tabular}{cp{10cm}c}
    \hline 
        \textbf{Database} & \makecell[l]{\textbf{Boolean Instructions}  }          \\ \hline
        ACM Digital Library & Title: communicat* AND (virtual OR vr) OR Abstract: communicat* AND (virtual OR vr) AND Full text: social  \\
        IEEE Xplore & ("Document Title": communicat* and (virtual or VR)) OR ("Abstract": communicat* and (virtual or VR)) AND ("Full Text Only": social)      \\ 
        Springer  & Title: communicat * AND (virtual OR vr) AND Full text: social   \\ \hline
    \end{tabular}
\end{table*}

  We used 'communicat*' to represent variations of the word "communication" such as communicate and communicating. The word 'virtual' was included as an exact search term, but we left the word 'reality' out to include different forms of expressing such settings and related technologies, interactions, user interfaces, and techniques, such as 'VR', and 'virtual environment'.

 We included the papers published from the past ten years, 2012 to 2022. The results were restricted to publications written in English. We chose to include only full papers because posters or adjunct publications often cannot provide a depth evaluation of users' communication behavior. This resulted in 1058 papers: 771 from ACM Digital Library, 246 from IEEE Xplore, 30 from Springer, and 11 from Google scholar. We compiled the titles and abstracts of these 1058 publications for screening in Phase 2.

\subsection{Screening and Eligibility}

We screened the title and abstract of 1058 papers in the screening phase using the following three criteria. 

\begin{enumerate}
\item \textbf{Communication.}
The paper must report an understanding of users' communication behavior. Interaction techniques such as facial reconstruction and haptic feedback can be used to facilitate communication. We excluded papers that do not focus on communication, such as understanding users' preferred collaborative activities in social VR. 

\item \textbf{User-to-user communication.}
The paper needs to focus on communication among users, including understanding users' communication behaviors in social VR or techniques to facilitate multi-user communication. We excluded papers that solve communication problems between a genuine user and a virtual agent.

\item \textbf{Immersive VR.} The user studies reported in the paper must be conducted in immersive VR environments with head-mounted displays. For the scope of this research, we excluded studies that explored augmented and mixed reality technologies, CAVEs, or other stereoscopic displays. We excluded papers aiming to facilitate communication between immersive VR users and outsiders. 

\end{enumerate}

Out of 1058, we included 76 papers for subsequent eligibility by screening the titles and abstracts using the inclusion criteria presented above.

In the eligibility phase, we then excluded 44 papers. Eighteen papers were excluded because they were not about immersive VR with HMD. 13 papers were excluded because they did not focus on communication between humans. Eight papers were excluded because they were not full papers (less than four pages). Five papers were excluded because they just mentioned communication in VR but did not dive into this topic. This step did not exclude papers based on the technologies used. Eventually, we included 32 papers for our detailed analysis.

\subsection{Data Set and Coding Process}
In this phase, we coded these 32 papers using thematic analysis \cite{thomas2008methods}. Two researchers first coded a randomly selected set of ten papers independently. The initial codebook was based on our research questions, such as communication approaches, research methods. Subsequently, the two researchers discussed their codes to gain a consensus on their understanding. Finally, two researchers used the consolidated codes to finish coding the remaining 22 papers.

We organized the codes into themes using affinity diagramming, which will be presented in the next section. We also conducted a quantitative analysis to show the statistics of relevant study designs. In the following two sections, we present our findings to answer the two research questions accordingly.

\section{Findings}
\subsection{Factors affecting communication behaviors (RQ1)}

We identified two key factors---\textit{Avatar and Non-verbal cues} that affect users' communication behaviors in social VR. According to the review, representation by avatars can facilitate communication for users by offering the sense of anonymity and social presence. Compared with video conferencing, the wide range of non-verbal cues in VR enrich communication for users by show user movements similar to reality (e.g., gesture, proximity, haptic feedback) or beyond reality (e.g., showing emojis above users' heads, or drawing and sketching in the mid-air).

\subsubsection{Avatar}

 Avatar-mediated communication (AMC) is a crucial component of the social VR experience. Avatar, as the digital representation of a human user in VR, enables users to communicate with one another over distance (e.g. the telegraph, the telephone, and the internet) via a natural user interface. 
  
 \textbf{Avatar-mediated communication facilitates communication.} Users can convey both verbal and non-verbal cues via their avatars while maintaining a level of anonymity and privacy~\cite{Maloney2020, BakerJ2021}. Abdullah et al. indicated that avatars can provide some sense of anonymity compared with video calls: the facial motion was reduced due to the lack of users' natural movement reconstruction \cite{Abdullah2021}. Baker et al. highlighted the importance of avatars playing a role in supporting shy or introverted older adults, who may be uncomfortable in traditional face-to-face settings, to feel more empowered to participate in social activities, and to avoid perpetuating negative aging stereotypes \cite{Baker2019, BakerJ2021}. Maloney's work showed that marginalized users employed non-verbal communication (e.g., muting themselves or using gestures, emojis, and body language) to protect themselves from unwanted behaviors while still being able to communicate in a comfortable and socially satisfying way without speaking \cite{Maloney2020}.
 
 Representation by avatar in social VR also allows users to feel co-present in virtual space, collapsing the physical and psychological distance that is often experienced when using other communication platforms\cite{Li2019, Zaman2015}. Smith's work explored how avatar can support communication by comparing embodied VR and no-embodied VR \cite{Smith2018}. With avatar as self-representation, their participants felt a much greater sense of social presence, stronger connection with their partner, more being willing to use body language to express themselves, more considerate and empathetic, less competitive and aggressive, and high communication quality.

 \textbf{Avatar's appearance affects the communication experience.} A high-fidelity avatar appearance benefits a comfortable collaboration and communication with the other users during and especially between tasks \cite{Hoppe2021}. Baker reported older adults' motivations for the avatar design process. The most commonly stated motivation was to create ‘realistic’ and personal likeness, and the rest of the motivations are relations, alter-egos, and playful/testing the software \cite{Baker2019}. Consistency of an avatar’s appearance and the ability to model a consistent virtual environment could help those with dementia to remember names and places as "they would always look the same". Furthermore, Baker and his colleague also revealed an avatar’s appearance can be adapted to suit a particular communication context. 
 Some research showed the possibility to design an emotional avatar based on bio-signal \cite{George2019, Bernal2017} to convey users' emotions more efficiently. Researchers used different visualizations to represent users' emotions, such as creating emotional beasts such \cite{Bernal2017} as self-representation, glowing avatar, avatar with an aura, colored arrows that indicate the pulse direction, and colored text. In this way, users can show emotional information that cannot be conveyed in the physical world, which can help users better express their emotions to one another.

 \textbf{The challenges of avatar-mediated communication.} There are two main concerning issues about this topic that can affect user communication quality in social VR: lacking facial cues and the error of body movement.
 The literature highlighted an immediate desire for the participants’ avatars to be more expressive, such as facial expressions, which give them a rich sense of communicating with another person \cite{Maloney2020, Tanenbaum2020, Baker2019, Semsioglu2021, Li2019}. In addition, lacking expressivity and behavioral anthropomorphism in avatars leads older adults to disassociate from them as a communication tool \cite{Baker2019}. However, it is still a challenge to reconstruct the avatar's facial expression according to the user's expression. To solve this problem, Vinnikov et al. proposed a prototype to enable facial expression while wearing a VR headset \cite{Vinnikov2017}, and Schwartz et al. reconstructed eye and face models for photorealistic facial animation \cite{Schwartz2020}. The error of body movement would create a negative experience \cite{Hoppe2021, Baker2019} for users when communicating in a virtual environment. It is worth noting that older people show particular sensitivity to involuntary movements caused by tracking errors \cite{Baker2019}, because they are particularly sensitive to social stereotypes that render the aging body as an object of disgust that makes them "liable to sanctions, both physical and symbolic". Therefore, it's important to reduce unnatural movements and interactions of avatars.

\subsubsection{Non-verbal communication}

Non-verbal communication plays a significant role in our daily lives since it provides extra emotions and meaningful interactions as a complement to verbal communication. As online social spaces evolve towards more natural embodied interaction, researchers have explored different types of non-verbal communication in social VR. In this section, we present the types of non-verbal interaction being explored, their outcomes, and the design implications for further improvement.

\begin{figure*}[]
\includegraphics[width=0.9\linewidth]{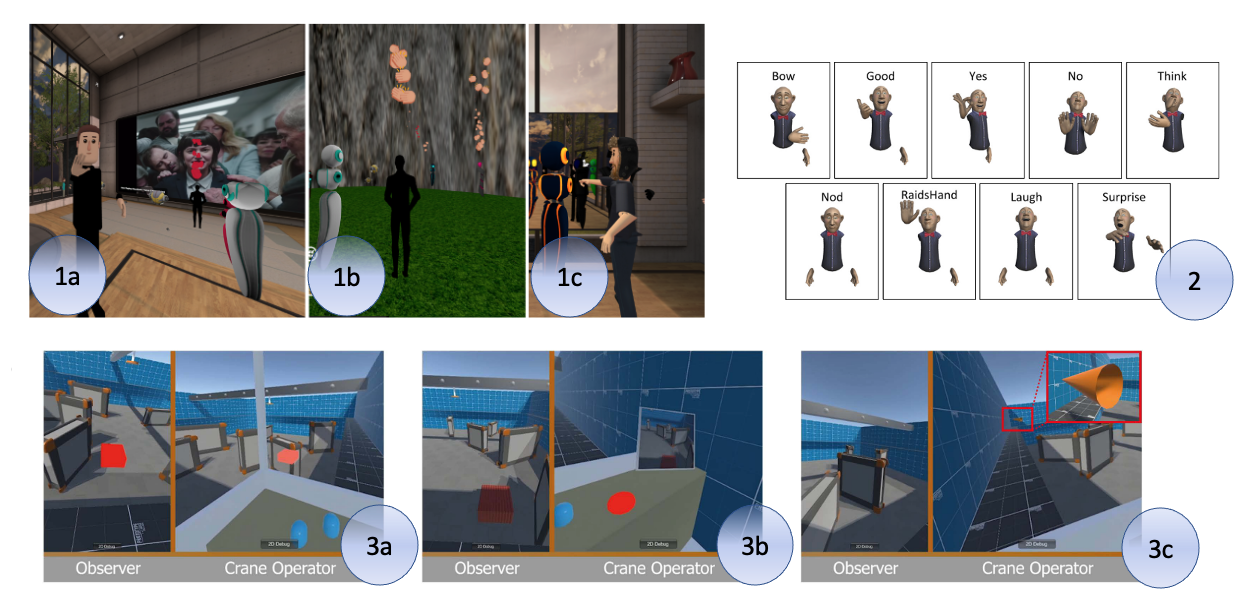}
\vspace{-0.2cm}
\caption{Examples of non-verbal behaviors in social VR. 1a, an extended hand, 1b, applause in the form of emotes, 1c, pointing \cite{Maloney2020}. 2, symbolic gesture condition \cite{Ide2020}. 3, visualization cues \cite{PhilippFreiwald2020}. 3a, object highlight, 3b, video mirror, 3c, view cone.} 
\label{fig:NVC}
\end{figure*}

\begin{table*}[htb!]
\renewcommand{\arraystretch}{1.3}
    \caption{Types of non-verbal interactions and their intentions}
    \label{tab:proscons}
    \Description{Table \label{tab:proscons} demonstrates the pros and cons of these three approaches.}
    \resizebox{\textwidth}{!}{
    \begin{tabular}{c|c|c|c|c|c|c|c|c}
    \hline
    \rowcolor[gray]{0.9} & \makecell[c]{Paying\\Attention} & Approving & \makecell[c]{Directing\\Attention} & Greeting & Provocating & Synchronization & \makecell[c]{Conveying\\Ideas} & \makecell[c]{Increasing \\Social Presence} \\
    \hline 
    \cellcolor[gray]{0.95} nodding behaviors \cite{Tanenbaum2020, Vindenes2021, Ide2020} & \checkmark & \checkmark & & & & & & \\
    \hline 
    \cellcolor[gray]{0.95}conversational turn \cite{Tanenbaum2020} & \checkmark & & & & & & & \\
    \hline 
    \cellcolor[gray]{0.95}eye gaze \cite{Tanenbaum2020, Vindenes2021} & \checkmark & & & & & & & \\
    \hline 
    \cellcolor[gray]{0.95}incorporated emojis \cite{Tanenbaum2020} &  & \checkmark & & & & & & \\
    \hline 
    \cellcolor[gray]{0.95}pointing \cite{Tanenbaum2020,Mei2021, Vindenes2021} &  & \checkmark & \checkmark & & & & & \\
    \hline 
    \cellcolor[gray]{0.95}patting one's own chest \cite{Tanenbaum2020} &  &  & \checkmark & & & & & \\
    \hline 
    \cellcolor[gray]{0.95}waving/dancing/kissing \cite{Tanenbaum2020, Mei2021} &  & \checkmark &  & \checkmark & & & & \\
    \hline 
    \cellcolor[gray]{0.95}pushing/poking/bumping \cite{Tanenbaum2020} &  &  &  &  & \checkmark & & & \\
    \hline 
    \cellcolor[gray]{0.95}lip/eye movement \cite{Maloney2020} &  &  &  &  & & \checkmark & & \\
    \hline
    \cellcolor[gray]{0.95}draw/sketch \cite{Mei2021, BakerJ2021, Men2018} &  &  &  &  & & & \checkmark & \\
    \hline
    \cellcolor[gray]{0.95}haptic \cite{Fermoselle2020} &  &  &  &  & & & & \checkmark \\
    \hline
    \cellcolor[gray]{0.95}\makecell[c]{communicative hand gestures\\(e.g., opening of palms)} \cite{Vindenes2021} &  &  &  &  & & & \checkmark & \\
    \hline
    \cellcolor[gray]{0.95}visual indicators \cite{PhilippFreiwald2020} &  &  & \checkmark &  & & &  & \\
    \hline
    \cellcolor[gray]{0.95}symbolic gestures(e.g., think) \cite{Ide2020} &  &  &  &  & & &  & \checkmark \\
    \hline
    \end{tabular}}
\end{table*}

Table ~\ref{tab:proscons} shows the types of non-verbal interactions and the corresponding intentions in social VR. Traditional modalities of non-verbal communication include facial expression, gaze, gesture, and proximity \cite{Tanenbaum2020}.

\textbf{Facial expression}: Facial expression communicate a person's emotional state and intention and allows for deliberate feedback such as a greeting smile and unintentional feedback, such as blushing or a reaction to a smell.

\textbf{Gaze}: Gaze can be used to communicate one's intention or emotion to others. Eye gaze can also increase the intensity of all facial emotions. 
In VR, eye gaze ("eye contact") is used to indicate one is paying attention \cite{Tanenbaum2020, Vindenes2021}.

\textbf{Gesture}: Gesture involves the movement of hands or other body parts. It is more vivid than plain language but may cause misunderstanding because gestures are learned and can vary across different cultures. Researchers have explored gestural behavior in VR extensively and we derive them into the following themes. 
1) Theme 1: As a similar form of communication to offline face-to-face interaction. This kind of gestural behavior includes nodding behaviors \cite{Tanenbaum2020, Vindenes2021, Ide2020}, conversation turn \cite{Tanenbaum2020}, pointing \cite{Tanenbaum2020,Mei2021, Vindenes2021}, patting one's own chest \cite{Tanenbaum2020}, waving/dancing/kissing for approving and greeting \cite{Tanenbaum2020, Mei2021}, communicative hand gestures (e.g., opening of palms) \cite{Vindenes2021}. People conduct the kind of gestural behavior to show that they are paying/directing attention, approving, and greeting. However, not all the gestural behavior are friendly. There is also interpersonal provocation that happened in VR such as poking and pushing \cite{Tanenbaum2020}. 
2) Theme 2: As an unconscious gestural behavior such as lip/eye movement which is synchronous with the user \cite{BakerJ2021}.
3) Theme 3: As a symbolic gesture to indicate people's intentions. For example, "nod and think" gestures (indicating the person is thinking), were found to be the most frequently used symbolic gestures in Ide et al's study \cite{Ide2020}. 

\textbf{Proximity}: Proximity refers to the distance between the avatars of the users as they interact with each other in the immersive VR~\cite{hall1966hidden}. 
Users can leverage their proximity to others to communicate anger, friendliness, and standoffishness through four distance zones: intimate, personal, social, and public. In immersive VR, how closely avatars stand together simulates spatial behavior in the offline world. People can control the distance by walking closer or further. Moreover, the audio cues of distance are enabled in social VR to inform users how close they are to other users, for example, hearing the sound of others' footsteps based on the distance \cite{Tanenbaum2020}.

\textbf{Haptic feedback}: Haptic feedback simulates users' communication and interaction in reality. Sugimori et al. proposed a novel avatar tracking controller with feedforward control that enables quick, accurate tracking and flexible motion in response to contacts between users. It frees avatar performers from the loads of performing as if contact occurred \cite{Sugimori2021}. In addition to simulate avatar motion, Fermoselle et al. added haptic interaction as one way to increase social presence \cite{Fermoselle2020}. For example, users can feel the touch when they give each other a "high-five" or pass documents among them.

\textbf{Other forms of non-verbal communication}: In addition to the above widely used forms of non-verbal communication, researchers have investigated other forms of non-verbal communication.  Tanenbaum et al. found that users incorporated \textit{emojis}, which appeared above their avatar's heads, to signal that they were on board with the direction of the conversation \cite{Tanenbaum2020}. \textit{Drawing or sketching in the mid-air} is also investigated to support users to convey ideas when collaborating \cite{Mei2021, BakerJ2021, Men2018}.
Philipp et al. investigated three distinct types of visual indicators including a 3D cone indicating the boundaries of a user’s field of view, highlighting the object a user is looking at, and displaying a direct video mirror of the user’s viewport \cite{PhilippFreiwald2020}. Hoppe et al. explored the social redirection technique that allows multi-user collaboration from a shared
perspective while at the same time providing a face-to-face interaction by shifting and modifying remote users’ avatars \cite{Hoppe2021}. 

\textbf{Non-verbal communication's outcomes.} Non-verbal communication can afford privacy and social comforts such as hiding voice data and other personally identifiable information and provide effective protection for marginalized users such as cis women, trans women, and disabled users \cite{Maloney2020}. Haptic or touch feedback can enhance the quality of the interaction experience by allowing users to exchange rich social cues in VR \cite{Fermoselle2020}.
Symbolic gestures (e.g., nod and think) and natural gestures (e.g., waving hands and pointing to objects) help users express their intentions and feelings and establish joint attention more easily to improve communication \cite{Ide2020}. Visual indicators of users’ perspectives can reduce the required verbal communication and therefore increase the efficiency of work within remote teams \cite{PhilippFreiwald2020}. Users can take a shared perspective to build a common understanding by the social redirection technique explored by Hoppe et al. and experience efficient collaboration and higher co-presence and the feeling of teamwork \cite{Hoppe2021}. Sketch in VR can help users express their design ideas quickly, and have a feeling of control, as well as a shared understanding of the design process and the final decisions with their collaborators \cite{Mei2021}. Specific prevented conversation scenes such as music composition, and sketch/annotations like ``signs and symbol'' make users feel positive and more effective to communicate during collaboration without interrupting or interfering with the music being created by collaborators \cite{Men2018}. Moreover, Farizi et al. explored using facial cues in VR for deception detection \cite{Farizi2019}.

 \subsection{Communication in Social VR v.s. Other forms of media (RQ1)}
 Surveyed papers compared communication behavior and experience among different mediums: VR, face-to-face, videoconference, and social media apps on the smartphone. Compared with other mediums, VR shows great potential in facilitating better communication. 
 
 Among these communication mediums, VR shows the most similarities to offline face-to-face \cite{Maloney2020}, in terms of spatial behavior, hand behavior, and facial expressions. Interactions in social VR blend the benefits of both digital communication and physical touch, which then provide an easier transition to initiate the social connection \cite{Tanenbaum2020}.
 Participants felt a great communication experience in VR. The user felt more excited and cheerful when they communicate in VR \cite{Li2019, Smith2018}. 
 VR provided a steady feeling of immersion \cite{Zaman2015, BakerJ2021, Tanenbaum2020, Tanenbaum2020}, thereby users would become more focused on the conversation \cite{Li2019}. The possible reason is VR blocks the external environment from entering the user's sight to let users feel less distracted by the environment \cite{Li2019}.
 In VR, participants perceived higher social presence than in videoconference and social media apps. People felt physically and emotionally closer in VR \cite{Li2019, Maloney2020} and showed the same connection with partners as in face-to-face communication \cite{Smith2018}. 
 
 With the avatar as self-representation, participants perceived a sense of anonymity \cite{BakerJ2021, Maloney2020, Abdullah2021} and felt relaxed \cite{Ide2020} when communicating in VR. VR may be able to play a role in supporting shy or introverted older adults \cite{BakerJ2021} and marginalized users \cite{Maloney2020}, who may be uncomfortable in traditional face-to-face settings, to feel more empowered to participate in social activities. VR showed effectiveness for the relaxation of tension by avatars \cite{Ide2020}, users showed a less formal interaction style when communicating in VR \cite{Abdullah2021}. 
 
 However, communication in VR also shows some drawbacks. The obvious problem is the uncomfortable caused by the heaviness and tightness of a head-mounted display (HMD), which keeps reminding participants that they are in VR \cite{Li2019}. Additionally, participants also indicated that it is inconvenient to communicate in VR because users need to wear an HMD and enter the virtual environment \cite{BakerJ2021}. Compared to social VR, the process of communication in a videoconference and social media app seems much easier. In addition, it was surprising that no papers reported that users felt cybersickness during the experiment.

\subsection{Evaluation methods for communication in social VR (RQ2)}

\subsubsection{Task Design}

 \textbf{Experimental Tasks}. There are different kinds of tasks to evaluate communication performance in different contexts (shown in table ~\ref{tab:studyTasks}). Communication often occurs in the process of cooperation, therefore, collaborative tasks \cite{Mei2021, Smith2018, PhilippFreiwald2020, Du2018, Liu2018, Zaman2015, Hoppe2021} were conducted to observe participants' communication behaviors. In addition, competitive task \cite{Liu2018}, independent task \cite{Liu2018, Vindenes2021}, and other kinds of tasks such as role play task \cite{Roth2018} and photo share task \cite{Li2019} were conducted to explore participants' behaviors in a different communication context. Some studies \cite{Maloney2020, Tanenbaum2020, Semsioglu2021, Men2018} asked participants to explore a social VR system with a given question in mind, while some studies \cite{Sugimori2021, Rungta2018, Farizi2019, Mayer2020, Ahmed2016, George2019, Dzardanova2021} invited participants to observe and score the prototype's performance, to evaluate how this prototype can facilitate the user's communication quality in a particular context. To observe user's communication behavior in a natural state, some studies let participants free talk with each other \cite{BakerJ2021, Baker2019, Fermoselle2020}, while some studies let participants talk with given topics  \cite{Ide2020, Seele2017, Vindenes2021, Miller2021, Smith2018}, such as intelligence problems, decision-making, negotiation, and allocation tasks, to evaluate user's behaviors and performance in different communication context in VR environment. 
 
 \begin{table}[htb!]
\renewcommand{\arraystretch}{1.3}
    \caption{Experimental tasks}    \label{tab:studyTasks}
    \begin{tabular}{cp{5cm}c}
    \hline
        \textbf{Number of each group}  & \textbf{Paper}          \\ \hline
        Collaborative task     & \cite{Mei2021, Smith2018, PhilippFreiwald2020, Du2018, Liu2018, Zaman2015, Hoppe2021}      \\
        Prototype scoring   & \cite{Sugimori2021, Rungta2018, Farizi2019, Mayer2020, Ahmed2016, George2019, Dzardanova2021} \\
        Prototype exploration     &  \cite{Maloney2020, Tanenbaum2020, Semsioglu2021, Men2018} \\
        Talking (given topics) &  \cite{Ide2020, Seele2017, Vindenes2021, Miller2021, Smith2018, Smith2018} \\
        Talking (freely)   & \cite{BakerJ2021, Baker2019, Fermoselle2020} \\
        Independent task   & \cite{Liu2018, Vindenes2021}\\
        Competitive task  & \cite{Liu2018}\\
        Role play task   & \cite{Roth2018}\\
        Photo share task   & \cite{Li2019}\\\hline
    \end{tabular}
    
\end{table}

  \textbf{Number of users per experimental group}. Communication usually takes place among several people. For our review papers (shown in table ~\ref{tab:userNum}), 65.2\% of them evaluated two-person communication performance in a VR environment, and 21.7\% of them reported the experience when three or four people communicate in a group. Furthermore, 17.4\% of papers conducted a user study with one participant each time, in which the user studies focused on evaluating the system or algorithm that facilitates multi-users communication, such as evaluating the effect of reverberation and spatialization on cocktail-party effect in multi-talker virtual environments \cite{Rungta2018}. However, no studies reported users' communication experiences in groups of more than five participants. Nowadays, many common VR scenarios, such as group meetings, include more than five users communicating simultaneously. Researcher needs to explore the challenges users may encounter when communicating with each other in these scenarios.

\begin{table}[htb!]
\renewcommand{\arraystretch}{1.3}
    \caption{Number of participants per experimental group}
    \label{tab:userNum}
    \begin{tabular}{cp{3cm}c}
    \hline
        \textbf{Number of participants}  & \textbf{\% of experiments  }          \\ \hline
        1     & 17.4\%      \\
        2    & 65.2\%      \\
        3 or 4     & 21.7\%  \\
        more the 5   & 0 \\ \hline
    \end{tabular}
\end{table}

 \textbf{Experimental site.} Among these papers, only two conducted the study with remote users \cite{BakerJ2021, Vindenes2021}. All other user experiments invite users to the laboratory for experiments. Users wearing HMD communicated in a VR environment in the same actual room, where they could even hear each other's voices in the real environment. It is questionable whether the communication behavior of users in the laboratory is natural. We believe that to obtain better user communication behavior and experience in the natural state, more user experiments must be tested among multiple remote users in the future.

\subsubsection{Data collection and analysis}

 To observe users' communication behaviors and evaluate users' communication quality, quantitative and qualitative methodology was conducted in surveyed studies. 
 
\begin{table}[htb!]
\renewcommand{\arraystretch}{1.3}
    \caption{Quantitative data and analysis method used in surveyed studies}
    \label{tab:QuanDataMethod}
    \begin{tabular}{cp{3.5cm}cp{3cm}}
    \hline
        \textbf{Quantitative data}   & \textbf{Paper }      \\ \hline
        System usability scale      & \cite{Sugimori2021, Mei2021, Seele2017, Roth2018,Fermoselle2020, Zaman2015, Ahmed2016, Hoppe2021, Schwartz2020, George2019}     \\
        Task performance & \cite{PhilippFreiwald2020, Du2018, Rungta2018, Farizi2019, Mayer2020, Roth2018, Hoppe2021, Dzardanova2021}  \\ 
        Social presence questionnaire  & \cite{Smith2018, Ide2020, Seele2017, Mayer2020, Roth2018,Li2019,Hoppe2021}   \\
        User experience      & \cite{Ide2020, Roth2018, Fermoselle2020, Li2019, Zaman2015, Ahmed2016, George2019} \\
        User behaviors     & \cite{Ide2020, Sugimori2021, Zaman2015, Miller2021, Abdullah2021, Liu2018}     \\
        Simulator sickness questionnaire   & \cite{PhilippFreiwald2020, Roth2018}  \\ 
        Raw NASA-Task Load Index     & \cite{Mayer2020, Hoppe2021}  \\
        Self-report questionnaire   & \cite{Roth2018, Seele2017}  \\ \hline
        \textbf{Quantitative analysis method} \\ \hline
        Scores means \& standard deviations      & \cite{Smith2018, Ide2020, Roth2018, Fermoselle2020, Hoppe2021, Hoppe2021, Abdullah2021, Dzardanova2021}  \\
        Analysis of Variance(ANOVA)   & \cite{PhilippFreiwald2020, Du2018, Rungta2018, Seele2017, Mayer2020, Roth2018, Ahmed2016}  \\
        Wilcoxon’s signed-rank test   & \cite{Sugimori2021, Li2019, Hoppe2021, Abdullah2021, Dzardanova2021}  \\ 
        Friedman test   & \cite{Ide2020, Li2019, Miller2021}  \\
        T-test   & \cite{Fermoselle2020,Miller2021,Abdullah2021}  \\
        Correspondence analysis   & \cite{Li2019}  \\
        Chi-square test   & \cite{Abdullah2021}  \\\hline
        
    \end{tabular}
\end{table}

 \textbf{Quantitative data}. Table ~\ref{tab:QuanDataMethod} shows the quantitative data and analysis method used in surveyed papers. Questionnaires are a frequently used approach to gathering quantitative data. Concerning communication system performance, the system usability scale \cite{Sugimori2021, Mei2021, Seele2017, Roth2018, Fermoselle2020, Zaman2015, Ahmed2016, Hoppe2021, Schwartz2020, George2019} is the most common subjective questionnaire used in surveyed studies aiming to investigate the performance of prototypes in promoting user communication. Furthermore, surveyed studies also collected task data to gain users' objective communication performance \cite{PhilippFreiwald2020, Du2018, Rungta2018, Farizi2019, Mayer2020, Roth2018, Hoppe2021, Dzardanova2021}, such as completion time, accuracy, answers to given questions, and so on. 
 For the user's subjective experience, a social presence questionnaire \cite{Smith2018, Ide2020, Seele2017, Mayer2020, Roth2018, Li2019, Hoppe2021} was used to investigate the user's sense of social participation and connection with their partner. Other user experience, such as user's emotion \cite{Ide2020, Roth2018, Fermoselle2020, Li2019, Zaman2015, Ahmed2016, George2019}, simulator sickness questionnaire (SSQ) \cite{PhilippFreiwald2020, Roth2018}, raw NASA-Task Load Index (raw TLX) \cite{Mayer2020, Hoppe2021} and self-report questionnaire (personality, technology expertise) \cite{Roth2018, Seele2017}, is used to investigate user's acceptance and feeling after communication in VR. To understand user behaviors some researchers \cite{Ide2020, Sugimori2021, Zaman2015, Miller2021, Abdullah2021, Liu2018} counted the number of gestures, talk duration, talk turn, and utterance frequency during communication and detected the user's bio-signal to gain subjective data for depth analysis. 

 To analyze those quantitative data, Wilcoxon’s signed-rank test \cite{Sugimori2021, Li2019, Hoppe2021, Abdullah2021, Dzardanova2021}, scores means, and standard deviations \cite{Smith2018, Ide2020, Roth2018, Fermoselle2020, Hoppe2021, Hoppe2021, Abdullah2021, Dzardanova2021}, Analysis of Variance(ANOVA) \cite{PhilippFreiwald2020, Du2018, Rungta2018, Seele2017, Mayer2020, Roth2018, Ahmed2016}, Friedman test \cite{Ide2020, Li2019, Miller2021}, t-test \cite{Fermoselle2020,Miller2021,Abdullah2021}, correspondence analysis \cite{Li2019}, chi-square test \cite{Abdullah2021} were used in surveyed papers.

\begin{table}[htb!]
\renewcommand{\arraystretch}{1.3}
    \caption{Qualitative data and analysis method used in surveyed studies}
    \label{tab:QuaDataMethod}
    \begin{tabular}{cp{3.5cm}cp{3cm}}
    \hline
        \textbf{Qualitative method}  & \textbf{Paper}          \\ \hline
        Observation and note  & \cite{Tanenbaum2020, Maloney2020, BakerJ2021, Smith2018, Baker2019,Semsioglu2021, Men2018,Li2019,Zaman2015,Hoppe2021,Miller2021}   \\
        Interviews      & \cite{Vindenes2021, Mei2021, Maloney2020, BakerJ2021, Smith2018, Du2018, Men2018, Li2019, Freeman2022}   \\
        Free feedback/comment & \cite{Ide2020, Roth2018, Fermoselle2020, Zaman2015, Hoppe2021, Vindenes2021, Miller2021}  \\
        Focus group     & \cite{Baker2019, Li2019, George2019}     \\
        Expert session     & \cite{Li2019}  \\ \hline
        \textbf{Quantitative analysis method} \\ \hline
        Thematic textual analysis   & \cite{Maloney2020,Tanenbaum2020,BakerJ2021,Mei2021,Smith2018,Baker2019,Semsioglu2021,Men2018,Li2019,Zaman2015,Vindenes2021,Freeman2022}  \\ 
        Context Mapping      & \cite{Li2019}  \\ \hline
        
    \end{tabular}
\end{table}

 \textbf{Qualitative data}. Table ~\ref{tab:QuaDataMethod} shows the qualitative data and analysis method used in surveyed papers. Regarding qualitative data collection, the most common method is to observe and note users' behaviors during the communication process \cite{Tanenbaum2020, Maloney2020, BakerJ2021, Smith2018, Baker2019,Semsioglu2021, Men2018,Li2019,Zaman2015,Hoppe2021,Miller2021}. The other common method is to conduct semi-structured or structured usability post-experience interviews \cite{Mei2021, Maloney2020, BakerJ2021, Smith2018, Du2018, Men2018, Li2019, Freeman2022, Vindenes2021} to explore user experience in-depth and discuss some possible future directions. Additionally, some researchers asked participants to give some free feedback/comments \cite{Ide2020, Roth2018, Fermoselle2020, Zaman2015, Hoppe2021, Vindenes2021, Miller2021} after the experiment. Focus group \cite{Baker2019, Li2019, George2019}, and expert sessions \cite{Li2019} were used in previous studies to understand better people's reactions and perceptions of certain experiences for future design.

 To analyze qualitative information, thematic textual analysis \cite{Maloney2020,Tanenbaum2020,BakerJ2021,Mei2021,Smith2018,Baker2019,Semsioglu2021,Men2018,Li2019,Zaman2015,Vindenes2021,Freeman2022} and context Mapping \cite{Li2019} was used in surveyed papers.

 It is worth noting that many papers regard users' verbal characteristics as a measure of users' communication quality and fluency, as well as user experience in different communication media, such as utterances, turn taking (i.e., how communication switches between different parties involved) \cite{Smith2018}, turn duration \cite{Li2019}, valence-arousal \cite{Liu2018, Semsioglu2021}, and the overlap of verbal communication \cite{Abdullah2021}.

\subsubsection{Task Implementation}

  \textbf{Head-Mounted Display}. Table ~\ref{tab:display} shows that Head-Mounted Display (HMD) was used in surveyed papers. HMD is used to render a VR environment and track users' information, which can impact user experience, information required and transferred, and the quality of communication. 29 of 32 papers reported VR HMD. Most of the research used Oculus Rift and HTC VIVE as experimental equipment. HMD can completely block external visual information from the real world, resulting in a more profound sense of immersion\cite{Zaman2015, BakerJ2021}. However, the heavy and tight VR HMD keeps reminding users that it is not the real world \cite{Li2019}.

\begin{table}[htb!]
\renewcommand{\arraystretch}{1.3}
    \caption{HMD used in surveyed studies}
    \label{tab:display}
    \begin{tabular}{cp{3cm}c}
    \hline
        \textbf{HMD} & \textbf{\% of studies  }          \\ \hline
        Oculus Rift      & 62.1 \%       \\
        HTC VIVE         & 37.9 \%  \\
        Other (Oculus Go, Samsung GearVR)   & 10.3\%  \\ \hline
    \end{tabular}
\end{table}

 \textbf{Other Tracker Devices}. Table ~\ref{tab:tracker} shows that other tracker devices were used in surveyed papers. User movement is a crucial component of non-verbal communication, including facial movement, body movement, hand gestures, user position, and orientation in the environment. Although HMD and hand controller can obtain the approximate position and orientation of the user's hands and head, these are far from enough to create a more expressive and realistic avatar. Therefore, 12 of 32 papers equipped exact trackers to require more information to reconstruct the user's facial information and body or gesture movements, while two papers detected the user's bio-signal to build an expressing avatar or evaluate the user's emotion. However, the lack of facial cues \cite{Maloney2020, Tanenbaum2020, Baker2019, Semsioglu2021, Li2019} and the error of avatar body movement are \cite{Baker2019, Hoppe2021} still reported by many works. Users desire a more expressive and natural avatar to convey their emotions when communicating with others in VR.

\begin{table}[htb!]
\renewcommand{\arraystretch}{1.3}
    \caption{Tracker hardware used in surveyed studies}
    \label{tab:tracker}
    \begin{tabular}{cp{3cm}cp{3cm}}
    \hline
        \textbf{Tracker} & \textbf{Proportion of the surveyed papers}            \\ \hline
        OptiTrack      &  3/32      \\
        Microsoft Kinect sensor   & 3/32  \\ 
        Leap Motion      &  2/32      \\
        Intel RealSense 3D Camera       & 2/32  \\
        In-depth Camera   & 2/32  \\ 
        Eye tracker       & 2/32  \\
        Bio-signal tracker   & 2/32  \\
        Other body tracker      & 1/32  \\\hline
    \end{tabular}
\end{table}

\subsubsection{Participants}

\textbf{Participants composition.}
 The studies had an average of 38.2 participants. The number of participants ranged from 10 to 210, with a mean of 38 and a median of 25. The age of participants ranged from 18 to 81. Among these participants, 61.69\% are male, 37.05\% are female, and 1.26\% are other. Some studies invited minority population as experimental participants. In the real world, minorities encounter more challenges than ordinary people in communication  \cite{Maloney2020}, such as harassment and discrimination. However, with an avatar's representation, VR brings a different perspective for minorities to communicate better. Among these surveyed papers, five papers reported older people communication experience in VR, in which two papers took older adults as targeted users to understand their communication needs in social VR \cite{Baker2019, BakerJ2021}. 4 papers reported non-cisgender users' experience, 2 papers reported the communication experience of disable people. The representation of an avatar in the VR world creates a sense of anonymity, which help minorities user avoid being treated as different. For example, Maloney et al.'s study \cite{Maloney2020} showed that non-cisgender users employed non-verbal communication (e.g., muting themselves or using gestures, emojis, body language) to protect themselves from unwanted behaviours but still communicate in a comfortable and socially satisfying means while not speaking. These users reported unique communication needs in these papers. However, no papers take these minority people (except the older adults) as their targeted users to comprehensively understand their communication needs. 
 
 \begin{table}[htb!]
 \renewcommand{\arraystretch}{1.3}
    \caption{Minority participants in surveyed studies}
    \label{tab:composition}
    \begin{tabular}{cp{3cm}c}
    \hline
        \textbf{Minority user} & \textbf{Paper}            \\ \hline
        Old people & \cite{Baker2019, BakerJ2021, Sugimori2021, Freeman2022, Maloney2020}  \\
        Non-cisgender & \cite{Freeman2022, Seele2017, Maloney2020}       \\ 
        Disable people  & \cite{Maloney2020, Freeman2022, BakerJ2021}     \\ \hline
    \end{tabular}
\end{table}
 
 \textbf{Relationship of participants.} Among these surveyed papers, only two papers \cite{Semsioglu2021, Li2019} recruited participants who have good relationships with each other, such as family members, friends or significant others. Most studies do not take the users who know each other as the research object to explore the user communication pattern or interaction design under certain relationships. In practice, communication often occurs between people who know each other in the real world. Future research needs to explore communication patterns and interaction design in different relationships.

\section{Discussion}

We conducted a systematic literature review to understand users' communication behaviors and experiences in social VR. In this section, we further discuss our key findings and propose six future research directions for researchers and designers.

\subsection{The uniqueness of immersive VR compared with other communication media}

 Avatar-mediated communication (AMC) in social VR can offer a number of compelling advantages. Avatar, as the digital representation of a human user in social VR, offers users different communication experiences from other media  \cite{Abdullah2021, Zaman2015, Li2019, Liu2018}. Avatar can provide anonymity and privacy to support shy or introverted people or marginalized users to better communicate in VR  \cite{BakerJ2021,Tanenbaum2020}. With the representation of an avatar, users feel a greater sense of social presence and stronger connection with their partner and are more willing to use body language to express themselves. Compared to communication in the video conferencing tools, people feel more considerate and empathetic, less competitive and aggressive, and have high communication quality \cite{Abdullah2021}.
 
 Additionally, non-verbal cues in social VR can be more varied than in face-to-face and video conferencing tools. For example, Tanenbaum et al. found that users incorporated emojis, which appeared above their avatar's heads, to signal that they were on board with the direction of the conversation such as using a thumb up to indicate an agreement \cite{Tanenbaum2020}. Drawing or sketching in the air was also investigated to support users to convey ideas when they were collaborating \cite{Mei2021, BakerJ2021, Men2018}. 

 However, the lacking of facial cues and the error of body movement of the avatar still negatively impact the user experiences in VR. Moreover, the heaviness, tightness, and inconvenience of wearing VR glasses are still a problem for daily use.

 \subsubsection{Future Direction 1: Explore more interaction methods to facilitate non-verbal communication.}
Baker et al. reported that using controllers to express emotions was not intuitive and frustrated participants when they accidentally pulled the wrong control \cite{BakerJ2021}. Combined with the suggestion of enabling alternative modes of control \cite{Maloney2020}, future work should investigate more flexible and intuitive interaction methods for the convenience of the general users and those with physical disabilities. 
 
 \subsubsection{Future Direction 2: Improve the recognition accuracy of facial expressions and body gestures. }
Although 12 of 32 papers equipped motion trackers to acquire information to reconstruct the user's facial information and body or gesture movement, many papers still reported the lack of facial cues and the error of avatar body movement. Participants desired a more expressive and natural avatar to convey their emotions when communicating with others in VR \cite{Maloney2020, Mei2021, Smith2018}. More accurate hand and finger tracking are needed to be explored, which would especially benefit deaf users by allowing them to use sign language \cite{Maloney2020}.

\subsection{Methodology to evaluate users' communication quality and experience}

 We investigated the methodology adopted in the reviewed papers, including experimental tasks, number of participants per experimental group, experimental site, data collection and analysis methods and targeted participants. We proposed four future directions through a comparative analysis of this information.

 \subsubsection{Future Direction 3: Explore the need of marginalized users.}
 Through the literature review, we found that few papers took minorities as their targeted users to gain a comprehensive understanding of their unique communication challenges and needs, such as children, non-cisgender, and people with hearing-impaired or speech-impaired. It is essential to understand their communication needs to make a more accessible society. As Tanenbaum et al. suggested that future work should investigate questions such as harassment, accessibility, and marginalized communities in social VR to create a socially relevant social VR experiences \cite{Tanenbaum2020}.

 \subsubsection{Future Direction 4: Explore the communication challenges and needs of participants with more diverse relationships.} 
 Among the reviewed papers, only two of them \cite{Semsioglu2021, Li2019} recruited participants who have a close relationship with each other, such as family members, friends or significant others. Most studies did not take the users who know each other as the research objects to explore their communication patterns or interaction design for certain relationships. In practice, communication often occurs between people who know each other in the real world. Future research needs to explore communication patterns and interaction design in different relationships.
 
 \subsubsection{Future Direction 5: Explore communication experience in a larger user group.} 
 Communication usually takes place among several people. In our reviewed papers, no studies reported users' communication experiences in groups of more than five participants. Many common VR scenarios, such as group meetings, included more than five users communicating simultaneously. Future work should explore potential challenges that social VR users may encounter when communicating with each other in these scenarios.

 \subsubsection{Future Direction 6: Explore communication behaviors and user experiences for remote users.} 
 Most of our reviewed papers invited users to a laboratory for experiments. Users were wearing HMD communicated in a VR environment in the same actual room, where they could hear each other's voices via the natural environment. It is questionable whether the communication behavior of users in the laboratory represents their behaviors in a remote setting, where they join the study in their familiar physical environments. We believe that to obtain better user communication behaviours and experiences in the natural state, more user experiments should be tested among multiple remote users in the future.



\section{Conclusion}


Communication experiences and quality are crucial in social VR for collaboration and building social connectedness. We have conducted a systematic review using the PRISMA method to understand how communication has been studied in social VR in the past ten years' published papers in the major digital libraries (i.e., ACM Digital Library, IEEE Xplore and Springer). According to our results, we identified the key factors that affect people's communication experience and quality in social VR, including the representation of avatar bringing the sense of anonymity and relax, various approaches to convey emotion and exchange information.
Furthermore, we investigated the methodology used in surveyed studies and identified research opportunities to improve the design of social VR applications to better support communication and make it more accessible in the social VR. We discussed and proposed future research directions according to the results of paper reviews, such as using social VR as communication media for remote family relatives.

\bibliographystyle{ACM-Reference-Format}

\renewcommand{\bibpreamble}{References marked with *  are in the set of reviewed papers.}

\bibliography{main}

\end{document}